\documentclass[10pt, conference, compsocconf]{IEEEtran}

\ifCLASSINFOpdf
  \usepackage[pdftex]{graphicx}
\else
  \usepackage[dvips]{graphicx}
\fi
\usepackage[cmex10]{amsmath}

\begin{document}
%
\title{Toward a Robust Diversity-Based Model to Detect Changes of Context}


\author{\IEEEauthorblockN{Sylvain Castagnos}
\IEEEauthorblockA{KIWI Team, LORIA\\
Campus Scientifique, B.P. 239\\
54506 Vand\oe{}uvre - France\\
sylvain.castagnos@loria.fr}
\and
\IEEEauthorblockN{Amaury L'Huillier}
\IEEEauthorblockA{KIWI Team, LORIA\\
Campus Scientifique, B.P. 239\\
54506 Vand\oe{}uvre - France\\
amaury.lhuillier@loria.fr}
\and
\IEEEauthorblockN{Anne Boyer}
\IEEEauthorblockA{KIWI Team, LORIA\\
Campus Scientifique, B.P. 239\\
54506 Vand\oe{}uvre - France\\
anne.boyer@loria.fr}
}


%


\maketitle

\begin{abstract}
Being able to automatically and quickly understand the user context during a session is a main issue for recommender systems. As a first step toward achieving that goal, we propose a model that observes in real time the diversity brought by each item relatively to a short sequence of consultations, corresponding to the recent user history. Our model has a complexity in constant time, and is generic since it can apply to any type of items within an online service (\textit{e.g.} profiles, products, music tracks) and any application domain (e-commerce, social network, music streaming), as long as we have partial item descriptions. The observation of the diversity level over time allows us to detect implicit changes. In the long term, we plan to characterize the context, \textit{i.e.} to find common features among a contiguous sub-sequence of items between two changes of context determined by our model. This will allow us to make context-aware and privacy-preserving recommendations, to explain them to users. As this is an on-going research, the first step consists here in studying the robustness of our model while detecting changes of context. In order to do so, we use a music corpus of 100 users and more than 210,000 consultations (number of songs played in the global history). We validate the relevancy of our detections by finding connections between changes of context and events, such as ends of session. Of course, these events are a subset of the possible changes of context, since there might be several contexts within a session. We altered the quality of our corpus in several manners, so as to test the performances of our model when confronted with sparsity and different types of items. The results show that our model is robust and constitutes a promising approach.
\end{abstract}

\begin{IEEEkeywords}
User Modeling; Diversity; Context; Real-Time Analysis of Navigation Path; Recommender Systems

\end{IEEEkeywords}

%
\IEEEpeerreviewmaketitle

\section{Introduction} \label{introduction} 
Despite their growing success in industry (e-commerce, social networks, VOD, music streaming platforms) and their impressive predictive performances~\cite{Simpson:2014}, two major user concerns frequently show up about recommender systems in online services. First, people are more and more preoccupied by privacy issues. To maintain a good trust level, we should thus provide models and algorithms that offer the best compromise between quality of recommendations, ethics as regards data collection~\cite{Cranor:2005}, and users' policy~\cite{Knijnenburg:2013}. Second, recommendations are still too often made out of context. Recommending is not only a question of maximizing the accuracy, but also providing relevant items at the right time in the good manner~\cite{Jones:2010}. This is the reason why the literature about context-aware recommender systems is increasing fast~\cite{Hariri:2014}.

Starting from these observations, we wondered what could possibly be the necessary and sufficient data to understand as quickly as possible the user context, and then to adapt recommendations. As regards privacy, Cranor suggests to favor methods where personal data are transient (\textit{i.e.} deleted after the task or the session)~\cite{Cranor:2005}. The system should also rely on item profiles, rather than user profiles. Thus, it is reasonable to study the short history of recently consulted items, and see what are the common features or differences that could explain or characterize the current user context. This line of reasoning implies that we have a precise description of each item available in the online service, or at least an exhaustive set of description attributes like those we have in product catalogs, but for every type of items (music tracks, social network profiles of users and companies, \ldots).

Besides these considerations, Castagnos~\textit{et al.} took an interest in the role of diversity within the user decision-making process~\cite{Castagnos:2010}. They provide two interesting conclusions within the frame of e-commerce applications. On one hand, the diversity in recommender systems seems to significantly improve user satisfaction, and is correlated to the intention to buy. On the other hand, the user need for diversity evolves over time, and should be carefully controlled to provide the correct amount of diversity and novelty. Bringing too much diversity risks to transform recommendations into novelty. Recent works confirmed that satisfaction is negatively dependent on novelty~\cite{Ekstrand:2014}, and badly-used diversity can lead users to mistrust the system~\cite{Castagnos:2013}. Finally, in~\cite{Castagnos:2010}, we showed that recommender systems should increase the diversity level at the end of a session to make users more confident in their buying decisions. Yet, predicting when the session will end is not an easy task.

This conclusion led us to ask if we could take the opposite view: would it be possible to monitor the diversity level within user sequences of consultations over time, and find connections between variations of diversity and changes of context? Through an exploratory research, we proposed the first model that measure the diversity brought by each consulted item, relatively to a short user history~\cite{LHuillier:2014}. We showed that variations of diversity often match with ends of session. However, these conclusions were made \textit{a posteriori}, \textit{i.e.} by analyzing the whole sequence of consultations for each user, and then knowing how each session ended and how the next session started. Furthermore, our model was built by considering that all consulted items were of the same type. As an example, if the active user is listening to music, it should be possible to measure the diversity between each pair of items.

In this paper, we want to bring this model a step further. First, we aim at investigating if it allows us to predict ends of session in real time, without knowing what happens next. Then, we will test the robustness of our model, by reconsidering our strong hypothesis according to which we always have a complete description of items. We will thus evaluate the performances of our model when we have sparse data about items. At last, we will extend our model to a situation where the active user consults different types of items (\textit{e.g.} music tracks, social network profiles, ...). In this case, it is not always possible to measure the diversity between items, since their attributes may be different. Thank to a corpus of more than 210,000 consultations, we show that the performances of our system remain stable up to 60\% of missing diversity measures.

The rest of this paper is organized as follows: Section~\ref{related-work} offers an overview of the literature as regards diversity and context in recommender systems. Section~\ref{model} is dedicated to the presentation of our model and our hypotheses about its robustness to sparsity and diversification of types of items. Section~\ref{experiment} presents and discusses its performances.

\section{Related Work} \label{related-work}
\subsection{Diversity in Recommender Systems} \label{diversity}
Diversity has long been proven to improve the interactions between users and recommender
systems~\cite{McGinty:2003}. This dimension is considered in two different ways in the literature. Some analyze the impact of diversity on users' behavior, while others integrate diversity in machine learning algorithms of recommender systems. 

Diversity has first been defined by Smyth and McClave~\cite{Smyth:2001}
as the opposite dimension to similarity. More precisely, this measure quantifies the
dissimilarity within a set of items. Thus, diversifying recommendations consists in determining the best set of items that are highly similar to the users' known preferences while reducing the similarity between those recommendations. A classification of diversity has been proposed by Adomavicius and Kwon~\cite{Adomavicius:2012}. It distinguishes individual diversity and aggregated diversity, depending on if we are interested in generating recommendations to individuals, or to groups of users. Here, we focus on individual diversity.

Many works focus on controlling the diversity level brought by recommender systems. Diversity was initially dedicated to content-based algorithms, especially in the case we have attribute values for each item. We distinguish 3 practices: we can compute the diversity between two items~\cite{Smyth:2001}, the diversity within a set of items~\cite{Ziegler:2005}, or the relative diversity brought by a single item relatively to a set of items~\cite{Smyth:2001} (see Equation~\ref{eq:reldiv}). These metrics have then been used in content-based filtering to reorder the recommendation list, according to a diversity
criterion~\cite{Bradley:2001, Zhang:2008}. In addition to these content-based algorithms, some works have focused on a way to integrate diversity in collaborative filtering~\cite{Ziegler:2005, Said:2012}.

In parallel to the integration of diversity in recommender systems, many user studies took interest in the role and perception of diversity. McGinty and Smyth showed that diversity improves the efficiency of recommendations~\cite{McGinty:2003}. Many works showed that diversity is perceived by users~\cite{Zhang:2008, Lathia:2010, Jones:2010}, and positively correlated to user satisfaction~\cite{Castagnos:2013, Ekstrand:2014}. Nevertheless, it came out that the user need for diversity evolves over time and diversity should not be integrated in the same way at each recommendation stage~\cite{McGinty:2003, Castagnos:2010}. At last, recent works focus on how the amount of diversity should be provided by recommender systems~\cite{Hasan:2014}.

Contrary to this literature, we do not want to adapt the amount of diversity in recommendations. We aim at observing the natural diversity level within users' navigation path to infer their context. Thus, the following subsection will be dedicated to this notion of context.

\subsection{Context-Aware Recommender Systems} \label{context}
Integrating the context into the recommendation process is an increasing research field known as \texttt{CARS}, acronym for Context Aware Recommender Systems. In their state-of-the-art, Adomavicius \textit{et al.} present several approaches like contextual modeling, pre/post filtering method for using contextual factors in order to adapt recommendation to the users' context~\cite{Adomavicius:2011b}. Contextual factors are all the information which can be gathered and used by a system to determine and characterize the current context of the user. For example, a system can use the location of the user to adapt the recommendation~\cite{Kaminskas:2013}. The most important drawbacks of these kinds of systems lies in the fact that they are invasive, by using personal informations and most often require a complex representational model. For example, such systems can use ontologies to determine user context~\cite{Chen:2014}. Yet, such an ontology cannot be transferred from one domain to another. As Adomavicius and Tuzhilin explain in their conclusion, ``most of the work on CARS has focused on the representation view of the context and the alternative methods have been underexplored''~\cite{Adomavicius:2011b}. This fact has also been highlighted by Hariri \textit{et al.} who have developed a \texttt{CARS} based on users' feedback on items presented in a interactive recommender system~\cite{Hariri:2014}. Even if this approach dynamically adapts to changes of context, it requires user effort to obtain user' feedback on which the system is based. We thus aim at proposing a similar method having the same objectives, but more transparent for users by relying on item profiles and users' navigation path.
In the following, we propose to distinguish two different types of context: explicit context and implicit context. Explicit context is close to the definition of contextual factors, that is to say physical context, social context, interaction media context and modal context are different kinds of explicit context~\cite{Adomavicius:2011b}. Conversely, implicit context will refer to the common characteristics shared by the consulted items during a certain time lapse. The motivation behind this notion is that detecting implicit context does not increase user involvement, enhances the privacy and can be used in any application domain without heavy modifications.

\section{Model and Hypotheses} \label{model}
\subsection{Overview} \label{overview}
\begin{figure}
\centering
\includegraphics[width=0.50\textwidth]{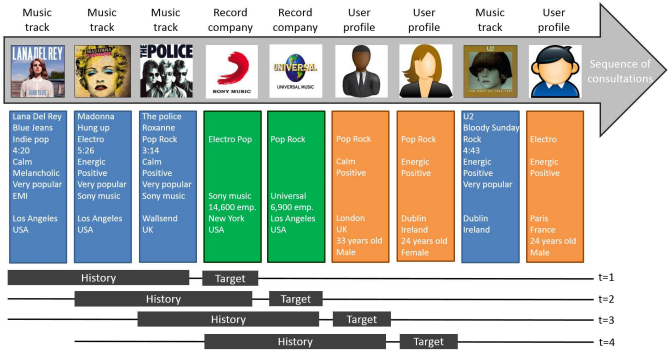}
\caption{Overview of our model.}
\label{fig:dance}
\end{figure}

As explained above, the role of our model is to monitor the diversity level within users' navigation path over time, and then derive their implicit context. Concretely, each time a user consults a new item, we compute the added value of this item -- called \texttt{target} -- relatively to a short history (\textit{i.e.} the $k$ previously consulted items) as regards to diversity. To provide a better understanding of our model, we will rely on an example shown in Figure~\ref{fig:dance}. Let us imagine an online service that allows users to listen to music, and to browse different kinds of profiles like we can do on social networks (profiles of other users, profiles of artists, information about record companies and so on). For each user, we can then pay attention to his/her sequence of consultations. In this example, we understand that there might be several contexts within a session, and several ways to classify them.


One strength of our model is that it allows us to measure in real time the diversity brought by each item, for each attribute independently, and for the whole set of attributes. Thus, it can be configured to detect and characterize various kinds of implicit contexts, or to cut the navigation path at some points where diversity reaches the highest levels (\textit{i.e.} what we called the changes of implicit context). In the rest of this article, we will give meaning to these changes of implicit context, by verifying that they match with some events such as ends of session in many cases. But, of course, there can be several successive implicit contexts, and several changes of context, within a session. Let us notice that, in the case where we want to force the detection of events and to optimize the characterization of the implicit context according to user's expectations, all we have to do is to complete a learning phase to find the optimal weight of each attribute within our computation of the diversity over time. The quality of our model has been demonstrated in~\cite{LHuillier:2014}. However, the purpose of this paper is to test the robustness of our model in the case where we have sparse data within item descriptions, that is to say detecting the same changes of implicit context with less data. We see two different scenarios which can explain sparse data. Either we have a single type of items (for example music tracks), but an incomplete description of each item, which is often the case in real applications. Or the users' navigation path are made of different types of items, and there may be a partial overlap of attributes between items. In Figure~\ref{fig:dance}, common attributes between items are displayed on the same line.

\subsection{Formalism} \label{formalism}
Before evaluating the robustness of our model, we will present it more formally and will introduce some notations. We call $U$~=\{\textit{$u_{1}$, $u_{2}$,..., $u_{n}$}\} the set of users. $u$ refers to the active user. $I$~=\{\textit{$i_{1}$, $i_{2}$,..., $i_{m}$}\} is the whole set of consulted items. The recent user history of size $k$ at time $t$, called $C_{k,t}^{u}$, can be written under the form of a sequence of items $<c_{t-k}^{u}$, ..., $c_{t-2}^{u}$, $c_{t-1}^{u}$, $c_{t-1}^{u}>$. At last, $A_i$~= \textit{$a_{1}$, $a_{2}$,..., $a_{h} $} is the set of attributes of an item $i$. Let us note that each consulted item, such as $c_{t}^{u}$, refers to an item $i$ of the set $I$.

Our model is a Markov model. At each time-step (\textit{i.e.} each time the active user consults a new item), our model computes the relative diversity brought by the new consulted item $c_{t}^{u}$ relatively to $C_{k,t}^{u}$. In order to do so, we strongly took inspiration from the formula proposed by Smyth and McClave~\cite{Smyth:2001} (see Equation~\ref{eq:reldiv}). The only difference here is that we count the number of times $s$ when we can compute the similarity between the target item $c_{t}^{u}$ and one of the items in the history $C_{k,t}^{u}$. As the active user can browse different types of items, there may be situations where there is no common attributes between two items, and no way to compute the similarity between this pair of items (\textit{i.e.} it returns NaN). Consequently, $s$ is included in $[0;k]$.
{\small
\begin{multline}\label{eq:reldiv}
\raggedright{
\scriptstyle RD(c_{t}^{u},C_{k,t}^{u})~=
\begin{cases}
 &\scriptstyle ~\text{NaN~if $C_{k,t}^{u}$~=~$\emptyset$ or if $s~=~0$,}\\
 &\scriptstyle ~\frac{\sum_{j=1..k}(1-sim(c_{t}^{u},c_{t-j}^{u}))}{s}~\text{otherwise.}
\end{cases}
}
\end{multline}
}

Measuring RD (Equation~\ref{eq:reldiv}) involves to compute the similarity between each pair of items, using Equation~\ref{eq:sim}. In this equation, the function $sim_{a}$ computes the similarity between two items relatively to a specific attribute~$a$. $\alpha_{a}$ is the weight of this attribute~$a$ in the computation of the similarity. In this paper, since we want mainly want to test the robustness of our model as regards sparse data, we will use a naive approach where each weight $\alpha_{a}$ is equal to 1. But we could parameter these weights to adapt our model, according to the kind of changes of implicit context and/or the kind of events we want to detect.
{\small
\begin{multline}\label{eq:sim}
\raggedright{
\scriptstyle sim(c_{t}^{u},c_{t-j}^{u})~=
\begin{cases}
 &\scriptstyle ~\text{NaN~if $(A_{c_{t}^{u}}\cap{}A_{c_{t-j}^{u}})$ or $c_{t}^{u}.a$ or $c_{t-j}^{u}.a~=~\emptyset$,}\\
 &\scriptstyle ~\frac{\sum_{a\in{}A_{c_{t}^{u}}\cap{}A_{c_{t-j}^{u}}} (\alpha_{a}~*~sim_{a}(c_{t}^{u},c_{t-j}^{u}))}{\sum_{a\in{}A_{c_{t}^{u}}\cap{}A_{c_{t-j}^{u}}} \alpha_{a}}~\text{otherwise.}
\end{cases}
}
\end{multline}
}

In Equation~\ref{eq:sim}, $i.a$ refers to the values (or set of values) of an attribute $a$ for a given item $i$. Starting from here, we developed 5 generic formulas to compute similarities per attribute, according to the type of attribute we have. If the values $i.a$ are expressed under the form of a list (\textit{e.g.} the attribute ``similar artists'' for a song), we will use Equation~\ref{eq:sima1}. 
\begin{equation}\label{eq:sima1}
sim_{a}(c_{t}^{u},c_{t-j}^{u})=\frac{card(c_{t}^{u}.a\cap c_{t-j}^{u}.a)}{min(card(c_{t}^{u}.a), card(c_{t-j}^{u}.a))}
\end{equation}

If the values $i.a$ correspond to intervals (\textit{e.g.} the attribute ``period of activity of a singer''), we will use Equation~\ref{eq:sima2}.
\begin{equation}\label{eq:sima2}
sim_{a}(c_{t}^{u},c_{t-j}^{u})=\frac{card(c_{t}^{u}.a\cap c_{t-j}^{u}.a)}{max(card(c_{t}^{u}.a), card(c_{t-j}^{u}.a))}
\end{equation}

If $i.a$ have binary values (\textit{e.g.} the mode of a song), we will use Equation~\ref{eq:sima3}. 
{\small
\begin{multline}
\raggedright{
\scriptstyle sim_{a}(c_{t}^{u},c_{t-j}^{u})~=
\begin{cases}
 &\scriptstyle ~1~\text{if}~c_{t-j}^{u}.a~=~c_{t}^{u}.a\text{,} \\
 &\scriptstyle ~0~\text{otherwise.}\hspace*{5em}
\end{cases}
\label{eq:sima3}}
\end{multline}
}

If $i.a$ take numerical values (\textit{e.g.} user ratings), we will use Equation~\ref{eq:sima4}.
\begin{equation}\label{eq:sima4}
sim_{a}(c_{t}^{u},c_{t-j}^{u})=e^{-10*\left(\frac{c_{t}^{u}.a-c_{t-j}^{u}.a}{max_{a} - min{a}}\right)^2}
\end{equation}

At last, if $i.a$ express coordinates (\textit{e.g.} the localization of two artists), we will use the Equation~\ref{eq:sima5}.
\begin{equation}\label{eq:sima5}
sim_{a}(c_{t}^{u},c_{t-j}^{u})=1~-~\frac{distance(c_{t}^{u},c_{t-j}^{u})}{max_{distance}}
\end{equation}

Finally, we are considering that there is a change of implicit context if the 4 conditions of Equation~\ref{eq:detection} are met. $\tau$ allows us to focus on relative diversity measures $RD(c_{t}^{u},C_{k,t}^{u})$ that exceed a given threshold.
\begin{multline}\label{eq:detection}
RD(c_{t-1}^{u},C_{k,t-1}^{u})<>\text{NaN}~\text{and}~RD(c_{t}^{u},C_{k,t}^{u})<>\text{NaN}\\
\text{and}~RD(c_{t-1}^{u},C_{k,t-1}^{u}) < RD(c_{t}^{u},C_{k,t}^{u})~\text{and}~RD(c_{t}^{u},C_{k,t}^{u}) > \tau
\end{multline}

\subsection{Hypotheses} \label{hypotheses}
The scientific question is now to test if our model is robust to a realistic situation where: (1) we do not know what will happen after the current time $t$, (2) we have sparse data as regards item descriptions. For these reasons, we will make 3 assumptions that will be discussed in Section~\ref{experiment}.
\noindent\fbox{
\begin{minipage}{7.95cm}
\textbf{H1.} {\it The extension of our model presented in this paper is able to detect changes of implicit context, without knowing the consultation at time $t+1$, and a large number of these detections match with events such as ends of session.}
\end{minipage}
}

This assumption has not been considered in our preliminary work in~\cite{LHuillier:2014}, since we were analyzing variations of diversity \textit{a posteriori} on the whole user's navigation path, knowing consultations at each time. We will thus check how many ends of session we can retrieve by only using data at time $t$, even if this does not lower the interests and relevancy of our other detections, as explained above (see Subsection~\ref{overview}).

\noindent\fbox{
\begin{minipage}{7.95cm}
\textbf{H2.} {\it The performances of our model remain stable when we reduce the amount of information available on items.}
\end{minipage}
}
 
Considering that we have a single type of items, we expect to retrieve the same amount of events and changes of implicit context.

\noindent\fbox{
\begin{minipage}{7.95cm}
\textbf{H3.} {\it The performances of our model remain stable when users consult different types of items.}
\end{minipage}
}

In this scenario, the attributes may be different from one type of items to another, leading to another form of sparsity.

\section{Experiments} \label{experiment}
In this section, we present 3 experiments we developed to validate these assumptions.

In the first experiment (\textbf{H1}), we test the ability of our model to detect changes of implicit context in real time, and check if the detected contexts could be correlated with some particular events like ends of sessions. However, unlike our exploratory research~\cite{LHuillier:2014}, our new model only uses data available at the current time $t$ (that is to say, we do not look at how diversity evolves beyond the current time). Indeed, our previous model was looking for local maxima on the curve of relative diversity and used thereby information unavailable at time $t$ to detect changes of context. In real situations, only present and past information are available. That is one of the reason which motivated us to extend our model (the other one is the consultations of different types of items). The principle of our model remains quite similar to~\cite{LHuillier:2014}. However, the inputs used to detect changes of context are different.

For each consulted item, we compute the corresponding values of relative diversity. As relative diversity can be computed for each attribute, there are as many relative diversity values as attributes. In this paper, we set the relative diversity of the current item to the average of all relative diversities per attribute. From now on, when we will talk about a relative diversity value according to an item, we will refer to the average relative diversity calculated from all the attributes for this item relatively to the history (Equation~\ref{eq:sim}). Inside a given context, we assume that the relative diversity of each item is quite constant and low, but that the relative diversity suddenly increases when changes of implicit context occur. This increase is due to the fact that different contexts do not share the same characteristics (\textit{i.e.} the same attribute values). Our model aims to detect these peaks of relative diversity over time. To achieve this, our model checks at each time-step if the conditions of Equation~\ref{eq:detection} are satisfied. In this case, we assume that $c_{t}^{u}$ is the first item of a new implicit context. For each new implicit context detected, we check if $c_{t}^{u}$ corresponds to the beginning of a new session.

In the second experiment (\textbf{H2}), we put to the proof our model by deleting information within our corpus. Indeed, data sparsity is a well-known problem in the field of recommender systems, and we want to know how our model can face this problem. In~\cite{LHuillier:2014}, we were using a complete dataset (\textit{i.e.} with no missing information about items), but that is rarely the case in real situations. For instance, in a musical corpus, we could have the song title and artist name for each track but some information like the release date, the popularity or the keywords may be missing. Thus, we want to test if:
\begin{itemize}
\item our model is able to compute a relative diversity value, even if some pieces of information about attributes are not known;
\item our model is robust to missing information and still performs well for detecting changes of context.
\end{itemize}

To answer these questions, we randomly delete values of attributes in our dataset, until we reach an intended rate of sparsity. We test the performances of our model for rates of sparsity between 1 and 99\%. Because of that random deletion, some similarity measures between two items, or even some relative diversity measures could not be computed. As soon as we can compute the similarity on at least one attribute for at least one pair of items (the target item and one of the items within the history), a value of relative diversity can be set for the target. 
Otherwise, if we cannot compute any similarity per attribute on any pair of items, we set the relative diversity of the target to \textit{NaN}. Let us notice that we set the diversity to \textit{NaN}, because a value of 0 would indicate that there is no diversity brought by the current item, not that the diversity cannot be calculated. Of course, we do not consider NaN values as changes of context (see Equation~\ref{eq:detection}).

In the last experiment (\textbf{H3}), the purpose is to examine the consequences of having several types of items in our dataset on context detection performances. Indeed, the previous experiments were tested with a single type of items but in practice, this may not be always the case. When the target item and the history items are of the same type (\textit{i.e.} music), the relative diversity can be computed on all attributes for all items (except when there are missing data). However, when these types may change from a consultation to another, the relative diversity can only be computed for common attributes (see Figure~\ref{fig:dance}).


Considering that our initial dataset contained a single type of items (songs), we modified it in order to test our third hypothesis. Criteria for simulating the different types of items were as follows: First, a number of types of items is determined, and all items are randomly assigned to a type of items. Afterward, for each type of items, we randomly select a subset of $x$ attributes (from the whole set of attributes) that will characterize these items. Another parameter, called $y$, corresponds to the minimum number of attributes in common with all the other types of items. Let us notice that the common attributes between pairs of types of items are not necessarily the same (\textit{i.e.} ($A_{type1}\cap{}A_{type2})<>(A_{type2}\cap{}A_{type3})$). In this way, we can artificially obtain a dataset composed of different types of items, with only a few attributes in common.

For instance, if the initial dataset contains 7 attributes ($A=\{a_{1},a_{2},a_{3},a_{4},a_{5},a_{6},a_{7}\}$) and we want to create 3 types with $x=4$ and $y=2$, we randomly get this kind of situation: $A_{type1}=\{a_{1},a_{4},a_{6},a_{7}\}$,  $A_{type2}=\{a_{1},a_{2},a_{3},a_{4}\}$, and $A_{type3}=\{a_{2},a_{3},a_{4},a_{6}\}$. In that case, $A^{type1}\cap{}A^{type2}=\{a_{1},a_{4}\}$, $A^{type1}\cap{}A^{type3}=\{a_{4},a_{6}\}$, and $A^{type2}\cap{}A^{type3}=\{a_{2},a_{3},a_{3}\}$.

\subsection{Material} \label{material}
In order to test our different hypotheses, we decided to base our evaluation on a musical dataset. This choice was made because musical items offer many advantages. First, musical items have their own consultation time, that is to say the time spent to consult a song cannot vary from a user to another. Second, meta data on songs can be easily retrieved using some specialized services like Echnonest\footnote{http://developer.echonest.com/} or Musicbrainz\footnote{https://musicbrainz.org/}. At last, users frequently listen to several songs consecutively, contrary to a movie corpus for example.

Our dataset contains 212,233 plays which were listened by 100 users. We obtained these consultations by using the Last.fm\footnote{http://www.lastfm.fr/} API to collect listening events from 28 June 2005 to 18 December 2014. Our dataset is made of 41,742 single tracks, performed by 5,370 single artists. In order to create the sessions for all the users, we assumed that a session is composed by a sequence of consultations without any interruption bigger than 15 minutes. When this threshold is reached, we consider that the user started a new session. According to this standard, we computed 22,212 sessions with an average of 9.6 consultations per session (42.71 min per session). Then, using the Echonest API, we gathered meta data on these songs. For each song, we retrieved 13 attributes: 7 of these attributes are specific to songs, and 6 of them are related to artists. 
\begin{itemize}
\item song attributes: duration, tempo, mode, hotttness, danceability, energy and loudness; 
\item artist attributes: hotttness, familiarity, similar artists (10 artists names), terms, years of activity, and location of the artist (geographical coordinates).
\end{itemize}

Table~\ref{tab:corpus} summarizes the values of the attributes. 
\begin{table*}[t]
\centering{
\scalebox{1.0}{
\begin{tabular}{|c||c|c|c|c|c|c|c||c|c|}
\hline 
 & \multicolumn{7}{c||}{Music} & \multicolumn{2}{c|}{Artist} \\ 
\hline 
Attribute & Duration & Tempo & Mode & Loudness & Energy & Hotttness & Danceability & Hotttness & Familiarity \\ 
\hline
Max & 4194 & 239 & 1 & 51.019 & 0.99 & 0.91584 & 0.9796 & 0.988956 & 0.912051 \\ 
\hline 
Min & 12 & 35 & 0 & 6.6479 & 0.00002 & 0.000782 & 0.039049 & 0.167657 & 0.136275 \\ 
\hline 
Average & 219.2446  & 128.1096 & 0.57353 & 40.72171 & 0.76148 & 0.334169 & 0.4331812 & 0.622260 & 0.631804 \\ 
\hline 
Deviation & 85.662929 & 30.30259 & 0.49456 & 4.19 & 0.2073564 & 0.12460 & 0.1679528 & 0.1233 & 0,1258315 \\ 
\hline 
\end{tabular}}}
\caption{Characteristics of the dataset}
\label{tab:corpus}
\end{table*}

\subsection{Results and Discussion} \label{results}
\textbf{Results as regards the first experiment (H1).} Previously, we presented Equation~\ref{eq:detection} which allow our model to determine if the current consultation is the start of a new implicit context. In order to fix the threshold $\tau$, we calculated the mean and the standard deviation of all values of relative diversity for all users within our corpus.

\begin{table}[h!]
\centering{
\begin{tabular}{|c|c|c|c|}
\hline 
   & Mean & Standard Deviation \\ 
\hline 
Relative diversity & 0.23 & 0.17 \\ 
\hline 
\end{tabular}}
\caption{Mean and Standard Deviation of RD}
\label{tab:statistique_rd}
\end{table}

In Table~\ref{tab:statistique_rd}, we can notice that the standard deviation is pretty high compared to the mean of the relative diversity. This result means that users' relative diversity over time takes a large range of values. We cannot know  \textit{a priori} the best value for $\tau$, since we do not know how many implicit contexts are present in our dataset. However, we previously assumed that diversity is pretty low within a given context and increases when a change of context occurs. This assessment can easily be confirmed \textit{a posteriori}, by noticing that the average level of relative diversity for consultations that correspond to a session opening ($average=0.36, standard deviation=0.13$) is much higher than those of other consultations ($average=0.21, standard deviation=0.16$). We finally decided to set $\tau$ to the global average of relative diversity within our dataset ($0.23$), so as to favor the detection of consultations above an average rate, but without fixing this threshold too high since there might be significant increase of diversity after a long period of decreasing (leading to values near the global average). When relative diversity exceeds this threshold and all the conditions of Equation~\ref{eq:detection} are satisfied, we consider that there is a change of implicit context. The results are reported in Table~\ref{tab:detection_naive}. 

\begin{table}[!h]
\centering{
\scalebox{0.8}{
\begin{tabular}{|c|c|c|c|}
\hline 
 & Existing & Detected by our model & Rate \\ 
\hline 
Sessions & 22,218 & 14,052 & 63.14 \% \\ 
\hline 
Implicit contexts & - & 37,743 & - \\ 
\hline 
\end{tabular}}
\caption{Performance of our model}
\label{tab:detection_naive}}
\end{table}

In total, our model detects 51,795 changes of implicit context. Among those changes of context, the number of sessions detected is important, since our model is able to detect more than 63\% of the sessions. This significant overlap between changes of context and events indicates that our model remains efficient when we only use information available at the current time (\textit{i.e.} without considering consultations at time $t+1$ and beyond), since we can easily justify/explain these changes of context by a end of session. This means that, when the explicit context changes (at least as regards the time dimension\footnote{Among other common explicit context factors such as localization, mood, people nearby and so on.} since there is a temporal gap between two sessions), the songs listened in those two explicit contexts do usually not share common characteristics (since they are in different implicit contexts).

We can also note that there are 37,743 changes of implicit context which do not match with changes of session. This is not a surprising result and can be explained in a simply manner. There can exist more than one implicit context within a session. We can easily imagine the case where a user starts listening to calm and down tempo songs, and suddenly changes to energetic and rapid tempo songs within the same session. 
As a conclusion of these results, we can say that our model seems to perform well by detecting possibly interesting points with the navigation path, that corresponds to changes of implicit context according to our definition, and can often by confirmed by changes of explicit context (events). But, as a perspective, we need to confront these results to real users, in order to study how they perceive and accept these implicit contexts, before using them as a support for recommender systems. Also, let us remind that we can easily change every parameter of our model (weights of attributes, size of history, value of the threshold $\tau$, ...) after a learning phase, to match users' expectations and maximize the acceptance and adoption rates.

\begin{figure}[!h]
\centering
\includegraphics[width=0.50\textwidth]{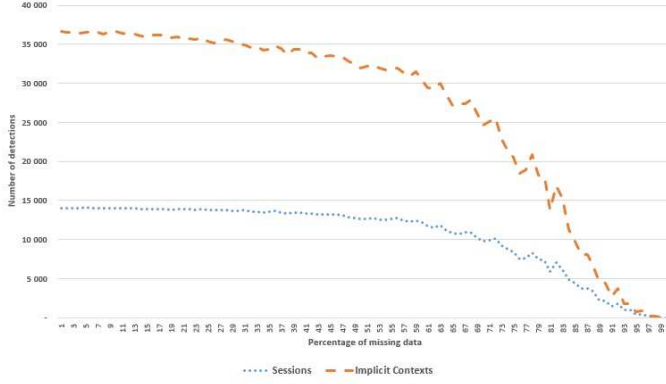}
\caption{Performance of our model against sparcity}
\label{fig:degradation_session}
\end{figure}

\textbf{Results as regards the second experiment (H2).} In order to understand how our model performs with a lack of data, we operated a controlled deterioration of our corpus. By controlled, we mean that the amount of missing data (that is to say missing values of attributes for the songs) was fixed for each execution. We monitored the number of sessions and implicit contexts detected, while progressively deteriorating the corpus percent after percent (see Figure~\ref{fig:degradation_session}).

From Figure~\ref{fig:degradation_session}, we can see that the performances of our model are pretty stable until up to 60\% of missing data. These results highlight the fact that our model can perform well, even with a large and realistic amount of missing data. 

\textbf{Results as regards the third experiment (H3).} Derived from some popular social networks like Facebook\footnote{https://www.facebook.com/}, LinkedIn\footnote{https://www.linkedin.com/}, or Yupeek\footnote{http://yupeek.com/}, we observed that the number of different types of items was usually around 4. That is why we decided to create 4 types of items from our initial corpus. On this basis, we tested different combinations as regards the number of attributes per item $x$ and the number of common attributes $y$. For each combination, we compute the number of sessions and implicit contexts detected. The results are presented in Table~\ref{tab:types_differents}. These values result from 10 executions, with the intent to limit bias due to the random selection of attributes. Indeed, according to the attributes which are selected for each type of items, the performance could vary as some attributes may be more representative than others in the detection of implicit contexts. 

\begin{table}[!h]
\centering
\scalebox{0.86}{
\begin{tabular}{|c|c||c|c||c|c|}
\hline 
\multicolumn{2}{|c||}{}& \multicolumn{2}{|c||}{Sessions (\%)} &\multicolumn{2}{|c|}{Implicit contexts (Number)} \\
\hline 
x & y & Avg. & $\sigma$(SD) & Avg. & $\sigma$(SD) \\ 
\hline
3 & 2 & 49.96 & 13.97&27563.6&6504.34 \\ 
\hline
4 & 2 & 52.80 & 7.54&29081.1& 3486.25\\ 
\hline
4 & 3 & 56.63 & 4.85&30695.1& 2053.30\\ 
\hline
5 & 2 & 59.07 & 2.61&32472.7& 1202.64\\ 
\hline
5 & 3 & 57.46 & 6.69&31618.3& 2817.34\\
\hline
5 & 4 & 59.17 & 5.59&32082.8& 2501.99\\
\hline
 6 & 2 & 57.25 & 5.24&31661.4& 2249.68\\
\hline
 6 & 3 & 56.31 & 6.42&31447.3& 2903.88\\
\hline
 6 & 4 & 57.54 & 7.40&31854.8& 3441.60\\
\hline
 6 & 5 & 59.76 & 4.16&32500.6& 1715.90\\
\hline
 7 & 2 & 60.10 & 1.92 &33594.9&984.30\\
\hline
 7 & 3 & 60.65 & 1.51&33983& 938.34\\
\hline
 7 & 4 & 59.62 & 3.16 &33156&1609.00\\
\hline
 7 & 5 & 59.53 & 4.43&32748.2&1815.43 \\
\hline
 7 & 6 & 59.94 & 3.89&33226.2& 1833.03\\
\hline
 8 & 2 & 60.43 & 1.73&33874.5& 1073.06\\
\hline
 8 & 3 & 60.84 & 1.31&34161.4& 795.48\\
\hline
 8 & 4 & 60.49 & 2.10&33844& 1404.76\\
\hline
 8 & 5 & 61.10 & 1.52&34320.6& 861.64\\
\hline
 8 & 6 & 61.20 & 2.52&33901.5&1277.09 \\
\hline
 8 & 7 & 61.65 & 1.36&33821.8& 931.85\\
\hline
 ... & ... & ... &... &...&... \\
\hline
 12 & 2 & 63.07 & 0.20 & 36242.9 & 407.87\\
\hline
 12 & 4 & 63.19 & 0.33 & 36322.4 & 481.87\\
\hline
 12 & 6 & 63.17 & 0.22 & 36494.8 & 348.32\\
 \hline
 12 & 8 & 63.00 & 0.21 & 36207.1 & 366.60\\
\hline
 12 & 10 & 63.10 & 0.21 & 36324.4 & 296.81\\
\hline
 13 & 63.246 & 63.246 & 0 & 36731 & 0\\
\hline
\end{tabular}}
\caption{{Performances of detection for different types of items}}
\label{tab:types_differents}
\end{table}

From Table~\ref{tab:types_differents}, we can observe that performances are quite good even if the number of attributes per type of items $x$ is low. Moreover, the highest the number of common attributes between types of items $y$ is, the more we detect changes of session and implicit contexts. We see that the standard deviation has high values when both the number of attributes $x$ and the number of common attributes $y$ are low. This confirms that all attributes have not the same impact in detecting changes of implicit context. It can be supposed that a difference between the value of the energy of two songs is more characteristic of a change of context than a variation of the artist location. Adapting the weight of each attribute in the calculation of the relative diversity for a given item is a perspective.

\section{Conclusions and Future Work} \label{conclusion}
Our model allows to monitor the natural diversity contained in users' navigation path over time and, although part of an on-going research, already presents many strengths to characterize user context. First, it has a complexity in constant time since, at each time-step, we only compute relative diversity on a fixed and small history size. This makes our model highly scalable. In addition, it preserves privacy, since it does not require personal information about the active user (even if it can make use of information that other users accept to share, as shown in Figure~\ref{fig:dance}) and allows to forget the navigation path beyond the recent history. At last, it is generic since our equations fit any kind of attributes, and does not require an ontology to put words on the context. 

One of the questions addressed in this paper was to check our ability to predict changes of implicit context at time $t$, without knowing what will happen next. So as to give meaning to these implicit contexts detected by our model, we tried to find a matching with explicit factors and events such as ends of session. Our results showed that we got a significant overlap between changes of implicit contexts and ends of session. Thus, this reinforce our conviction that this model highlights interesting points within users' navigation path. First, it allows us to anticipate ends of session, and will then be useful to adapt recommendations when users are near to reach a decision. Second, the changes of implicit context detected by our model that do not match with events are very promising results to be, on the long-term, able to formally characterize the user context and provide context-aware recommendations that fit privacy issues. 
Another purpose of this paper was to test the robustness of our model when confronted to sparse data. We distinguished two different scenarios where we have a single type of items with incomplete descriptions, or several types of items with small intersections of attributes. In both cases, the performances of our model remained stable in tough conditions, with about 60\% of missing data.

Among our perspectives, we aim at confronting our model to real users, so as to measure their perception and acceptance rate of implicit contexts. We expect to map implicit and explicit contexts so as to reach the same performances as systems based on explicit contexts, but with a deeper consideration of privacy issues. Finally, by characterizing implicit contexts, we will be able to explain recommendations based on implicit contexts and provide new interaction modes to make user decisions easier. 

\section*{Acknowledgements} 
This work was financed by the region of Lorraine and the Urban Community of Greater Nancy, in collaboration with the Yupeek company.





%

\end{document}